\documentclass[12pt]{article}

\usepackage{aasms4,tighten,natbib,psfig}

\def\la{\mathrel{\hbox{\rlap{\hbox{\lower4pt\hbox{$\sim$}}}\hbox{$<$}}}}
\def\ga{\mathrel{\hbox{\rlap{\hbox{\lower4pt\hbox{$\sim$}}}\hbox{$>$}}}}

\newcommand{\be}{\begin{eqnarray}}
\newcommand{\ee}{\end{eqnarray}}

\newcommand{\msol}{\ifmmode{{\rm M}_\odot}\else{M$_\odot$}\fi}
\newcommand{\foe}{\ifmmode{10^{51}}\else{$10^{51}$}\fi}

\newcommand{\xni}{\ifmmode{{\rm X}_{\rm Ni}}\else{X$_{\rm Ni}$}\fi}
\def\ang{\hbox{\AA}}
\def\Teff{\ifmmode{T_{\rm eff}}\else{\hbox{$T_{\rm eff}$} }\fi}
\def\Rzero{\ifmmode{R_0}\else{\hbox{$R_0$} }\fi}

\def\HPJ{{\tt HP J200}}
\def\SP2{{\tt IBM SP2}}
\def\GCel{{\tt GCel}}
\def\GCpp{{\tt GC/PP}}
\def\MPI{{\tt MPI}}
\def\HPF{{\tt HPF}}
\def\PC2{{\tt PC$^2$}}
\def\piofs{{\tt PIOFS}}
\def\RT{radiative transfer}
\def\RTC{radiative transfer code}
\def\nrt{N_{\rm RT}}
\def\nnode{N_{\rm node}}
\def\logg{\log(g)}

\def\inu{\ifmmode{I_{\nu}}\else{\hbox{$I_{\nu}$} }\fi}
\def\snu{\ifmmode{S_{\nu}}\else{\hbox{$S_{\nu}$} }\fi}
\def\jnu{\ifmmode{J_{\nu}}\else{\hbox{$J_{\nu}$} }\fi}

\def\fep{\ifmmode{{\rm Fe II}}\else\hbox{Fe~II }\fi}

  at 12.0 true pt 

\def\phoenix{{\tt PHOENIX}}

\def\phoenix{{\tt PHOENIX}}

\def\b{\beta}

\def\L{\Lambda}

\def\rout{\ifmmode{r_{\rm out}}\else\hbox{$r_{\rm out}$}\fi}
\def\tmax{\ifmmode{\tau_{\rm max}}\else\hbox{$\tau_{\rm max}$}\fi}
\def\tstd{\ifmmode{\tau_{\rm std}}\else\hbox{$\tau_{\rm std}$}\fi}
\def\vmax{\ifmmode{v_{\rm max}}\else\hbox{$v_{\rm max}$}\fi}
\def\muE{\ifmmode{\mu_{\rm E}}\else\hbox{$\mu_{\rm E}$}\fi} 
\def\pE{\ifmmode{p_{\rm E}}\else\hbox{$p_{\rm E}$}\fi} 
\def\bmax{\ifmmode{\b_{\rm max}}\else\hbox{$\b_{\rm max}$}\fi}

\def\ang{\hbox{\AA}}

\def\Teff{\hbox{$\,T_{\rm eff}$} }

\def\rout{\hbox{$r_{\rm out}$} }

\def\chistd{\ifmmode{\chi_{\rm std}}\else\hbox{$\chi_{\rm std}$}\fi}

\def\k{\,{\rm K}}
\def\msol{$M_\odot$}
\def\foe{10^{51}}

\def\xni{{\rm X}_{\rm Ni}}

\def\lstar{\ifmmode{\Lambda^*}\else\hbox{$\Lambda^*$}\fi} 
\def\Rop{\ifmmode{[R_{ij}]}\else\hbox{$[R_{ij}]$}\fi}

\def\Rji{\ifmmode{[R_{ji}]}\else\hbox{$[R_{ji}]$}\fi}
\def\Rstar{\ifmmode{[R_{ij}^*]}\else\hbox{$[R_{ij}^*]$}\fi}

\def\Rjistar{\ifmmode{[R_{ji}^*]}\else\hbox{$[R_{ji}^*]$}\fi}
\def\DRji{\ifmmode{[\Delta R_{ji}]}\else\hbox{$[\Delta R_{ji}]$}\fi}
\def\DRij{\ifmmode{[\Delta R_{ij}]}\else\hbox{$[\Delta R_{ij}]$}\fi}

\def\ns{\ifmmode{N_{\rm s}}          
        \else\hbox{$N_{\rm s}$}\fi}

\def\mat#1{{\bf #1}}     
\def\vek#1{{#1}}         

\newcount\eqcount
\def
  \nummer{
    \global\advance\eqcount by 1
    (\the\eqcount)
  }

\def
  \numadv{
    \global\advance\eqcount by 1
  }

\def
   \numout#1{
     (\the\eqcount #1)
  }

\def\ivek#1#2{\ifmmode{\vek{I}^{#1}_{#2}}
        \else\hbox{$\vek{I}^{#1}_{#2}$}\fi}

\def\tmat#1#2{\ifmmode{\mat{t}^{#1}_{#2}}
        \else\hbox{$\mat{t}^{#1}_{#2}$}\fi}
\def\rmat#1#2{\ifmmode{\mat{r}^{#1}_{#2}}
        \else\hbox{$\mat{r}^{#1}_{#2}$}\fi}
\def\bvek#1#2{\ifmmode{\beta^{#1}_{#2}}
        \else\hbox{$\beta^{#1}_{#2}$}\fi}

\def\lp{\ifmmode{\lambda^+_\tau}           
        \else\hbox{$\lambda^+_\tau$}\fi}
\def\lm{\ifmmode\lambda^-_\tau             
        \else\hbox{$\lambda^-_\tau$}\fi}

\def\la{\mathrel{\hbox{\rlap{\hbox{\lower4pt\hbox{$\sim$}}}\hbox{$<$}}}}
\def\ga{\mathrel{\hbox{\rlap{\hbox{\lower4pt\hbox{$\sim$}}}\hbox{$>$}}}}

\begin{document}
\bibliographystyle{natbib-apj}

\title{Parallel Implementation of the {\tt PHOENIX} Generalized Stellar
Atmosphere Program}
\author{Peter H. Hauschildt\altaffilmark{1,2}, E.~Baron\altaffilmark{3},
and France Allard\altaffilmark{4}}
 
\altaffiltext{1}{Dept. of Physics and Astronomy, 
University of Georgia, Athens, GA 30602-2451; yeti@hal.physast.uga.edu}

\altaffiltext{2}{Dept. of Physics and Astronomy, Arizona State
University, Tempe, AZ 85287-1504}

\altaffiltext{3}{Dept. of Physics and Astronomy, University of
Oklahoma, 440 W.  Brooks, Rm 131, Norman, OK 73019-0225;
baron@phyast.nhn.ou.edu}

\altaffiltext{4}{Dept. of Physics, Wichita State University,
Wichita, KS 67260-0032;
allard@eureka.physics.twsu.edu}

\begin{abstract}
We describe the parallel implementation of our generalized stellar
atmosphere and NLTE radiative transfer computer program \phoenix. We
discuss the parallel algorithms we have developed for radiative
transfer, spectral line opacity, and NLTE opacity and rate calculations.
Our implementation uses a MIMD design based on a relatively small
number of \MPI\ library calls. We report the results of test 
calculations on a number of different parallel computers and
discuss the results of scalability tests.
\end{abstract}

\section{Introduction}


Much of the astrophysical information that we possess has been obtained
via spectroscopy. Spectroscopy of astrophysical objects allows us to
ascertain velocities, temperatures, abundances, sizes, and luminosities
of astrophysical objects. Much can be learned from some objects just by
examining line ratios and redshifts, but to really understand an
observed spectrum in detail often requires detailed fits to synthetic
spectra. Detailed synthetic spectra also serve to test theoretical
models for various astrophysical objects. Since many astrophysical
objects of interest feature relativistic flows (e.g., supernovae, novae,
and accretion disks), low electron densities (e.g., supernovae and
novae), and/or molecular formation (e.g., cool stars, brown dwarfs, and
giant planets), detailed models must include special relativistic effects and
a very large number of ions and molecules in the equation of state in
order to make quantitative predictions about, e.g., abundances. In
addition, deviations from local thermodynamic equilibrium (LTE) must be
considered to correctly describe the transfer of radiation and the
emergent spectrum.

We have developed the spherically symmetric special relativistic non-LTE
generalized radiative transfer and stellar atmosphere computer code {\tt
PHOENIX} \cite[]{phhs392,phhcas93,hbfe295,phhnov95,faphh95,phhnovfe296,snefe296} which
can handle very large model atoms as well as line blanketing by millions
of atomic and molecular lines.  This code is designed to be very
flexible, it is used to compute model atmospheres and synthetic spectra
for, e.g., novae, supernovae, M and brown dwarfs, white dwarfs and
accretion disks in Active Galactic Nuclei (AGN); and it is highly
portable. When we include a large number of line transitions, the line
profiles must be resolved in the co-moving (Lagrangian) frame. This
requires many wavelength points, (we typically use 50,000 to
150,000 wavelength points).  Since the CPU time scales 
linearly with the number of wavelength points the CPU time requirements
of such a calculation are large. In addition, each NLTE radiative
rate for both
line and continuum transitions must be calculated and stored at every
spatial grid point which requires large amounts of storage and
can cause significant performance degradation if the corresponding
routines are not optimally coded.


In order to take advantage of the enormous computing power and vast
memory sizes of modern parallel supercomputers, both potentially
allowing much faster model construction as well as more detailed models,
we have implemented a parallel version of {\tt PHOENIX}. Since the code
uses a modular design, we have implemented different parallelization
strategies for different modules in order to maximize the total parallel
speed-up of the code. In addition, our implementation allows us to change
the load distribution onto different nodes both via input files and
dynamically during a model run, which gives a high degree of flexibility
to optimize the performance on a number of different machines and for a
number of different model parameters.

Since we have both large CPU and memory requirements we have initially
implemented the parallel version of the code on the \SP2\ using the
\MPI\ message passing library \cite[]{mpistd}.  Since the single
processor speed of the \SP2\ is high, even a small number of
additional processors can lead to significant speed-up. We have chosen
to work with the \MPI\ message passing interface, since it is both
portable \cite[public domain implementations of \MPI\ are readily
available cf.~][]{mpich}, running on dedicated parallel machines and
heterogeneous workstation clusters and it is available for both
distributed and shared memory architectures.  For our application, the
distributed memory model is in fact easier to code than a shared
memory model, since then we do not have to worry about locks and
synchronization, etc.\ on {\em small} scales and we, in addition,
retain full control over interprocess communication.  This is
especially clear once one realizes that it is fine to execute the same
code on many nodes as long as it is not too CPU intensive, and 
avoids costly communication,

 An alternative to an implementation with \MPI\ is an implementation 
using High Performance Fortran (\HPF) directives (in fact, both
can co-exist to improve performance). However, the process of 
automatic parallelization guided by the \HPF\ directives is presently
not yet generating optimal results because  the 
compiler technology is still very new. In addition, \HPF\ compilers are not
yet widely available and they are currently not available for 
heterogeneous workstation clusters. \HPF\ is also more suited for
problems that are purely data-parallel (SIMD problems)
and would not benefit much from a MIMD approach. An optimal
\HPF\ implementation of \phoenix\ would also require a significant
number of code changes in order to explicitly instruct the compiler
not to generate too many communication requests, which would slow
down the code significantly. The \MPI\ implementation requires
only the addition of a  few explicit communication requests, which can be 
done with a small number of library calls.

\section{Basic numerical methods}

%
The co-moving frame radiative transfer equation for spherically
symmetric flows can be written 
 as \cite[cf.~][]{found84}:

\vbox{\be
&\quad&\gamma (1+\beta\mu)\frac{\partial\inu}{\partial t} + \gamma (\mu +
\beta) \frac{\partial\inu}{\partial r}\nonumber\\
& +& \frac{\partial}{\partial
\mu}\left\{ \gamma (1-\mu^2)\left[ \frac{1+\beta\mu}{r}
\right.\right.\nonumber\\
&\quad&\left.\left. \quad -\gamma^2(\mu+\beta)
\frac{\partial\beta}{\partial r} -  
\gamma^2(1+\beta\mu) \frac{\partial\beta}{\partial
t}\right] \inu\right\} \nonumber\\
&-&  \frac{\partial}{\partial
\nu}\left\{ \gamma\nu\left[ \frac{\beta(1-\mu^2)}{r}
+\gamma^2\mu(\mu+\beta) \frac{\partial\beta}{\partial r}
\right.\right.\nonumber\\
&\quad& \left.\left.\quad  +
\gamma^2\mu(1+\beta\mu) \frac{\partial\beta}{\partial
t}\right]\inu\right\}\label{fullrte}\\
&+&\gamma\left\{\frac{2\mu+\beta(3-\mu^2)}{r}\right.
\nonumber\\
&\quad&\quad
\left. +\gamma^2(1+\mu^2+2\beta\mu)\frac{\partial\beta}{\partial r} + 
\gamma^2[2\mu + \beta(1+\mu^2)]\frac{\partial\beta}{\partial
t}\right\}\inu \nonumber\\
&\quad& = \eta_\nu - \chi_\nu\inu.\nonumber
\ee}
\noindent
We set  $c=1$; $\beta$ is the velocity; and
$\gamma = (1-\beta^2)^{-1/2}$ is the usual Lorentz factor. 
Equation~\ref{fullrte} is a integro-differential equation, since the
emissivity $\eta_\nu$ contains \jnu\unskip, the zeroth angular moment of
\inu\unskip:
\[ \eta_\nu = \kappa_\nu \snu + \sigma_\nu \jnu, \]
 and 
\[ \jnu = 1/2 \int_{-1}^{1} d\mu\, \inu, \]
where $\snu$ is the source function, $\kappa_\nu$ is the absorption
opacity, and $\sigma_\nu$ is the scattering opacity.
With
the assumption of time-independence $\frac{\partial\inu}{\partial t} =
0$ and a monotonic velocity field Eq.~\ref{fullrte} becomes a boundary-value problem in
the spatial coordinate and an initial value problem in the frequency
or wavelength coordinate. The equation can be written in operator form
as:
\be
\jnu = \L_\nu \snu,
\ee
where $\L$ is the lambda-operator. 

\section{Machines used for testing}

 We were able to test the algorithms described here on a few
different machines. In this section we describe briefly the 
characteristics of each system for reference. There are significant
differences between the architectures of the various systems, so we
could test the behavior of the code under very different conditions.

\subsection{\SP2}
The \SP2\ is a distributed memory machine based on the {\tt IBM
Power2} chipset. Each node is, effectively, an independent workstation
with its own operating system, paging space, and local disk space. The
nodes are connected in a flat topology, with a high-speed crossbar
switch that has a peak performance of $30\,$MB/s. We ran tests on the
\SP2's of the University of Oklahoma and the Cornell Theory Center
equipped with $62.5\,$MHz ``thin-node 2'' nodes with 128 to $512\,$MB
memory, $\ge 1\,$GB local disk space. In addition, both \SP2's have a
parallel filesystem (\piofs) that allows parallel IO to improve
performance.

\subsection{\HPJ}
We were able to run a number of tests on a dual processor \HPJ. This
machine is a SMP design with two {\tt HP} {\tt PA-7200} processors
running at $100\,$MHz. Our test machine has 128MB of memory and
a fast 3GB scratch filesystem. We use the public domain \MPI\ 
implementation {\tt MPICH} \cite[]{mpich} compiled on this machine and 
the {\tt HP} Fortran
compiler and libraries for the test calculations.

\subsection{\tt Parsytec GigaClusters}
We have run tests of the \RTC\ on two {\tt Par\-sytec} {\tt  Giga\-Clusters}
of the Paderborn Center for Parallel Computing (\PC2). We could not yet
run tests with \phoenix\ because of memory and disk limitations on these
machines, but the \RT\ calculations give an indication on the performance
and scalability of these systems.

\subsubsection{\GCpp}
The {\tt Parsytec} {\tt GigaCluster} {\tt Power} {\tt Plus} (\GCpp) uses
PowerPC 601 CPU's at a clockspeed of $80\,$MHz. Each processor has a
peak performance of $80\,$MFlops and a {\tt LINPACK 100x100} performance
of $15\,$MFlops.  The machine of the \PC2\ has $64\,$MB memory per
processing element (2 CPU's). The communication between processors has a
peak performance of $20\,$MB/s, but the sustained performance is only
$3.3\,$MB/s. The \GCpp\ runs under {\tt PARIX 1.3.1} and uses the {\tt
Motorola} Fortran compiler and \PC2's version of the \MPI\ libraries.

\subsubsection{\GCel}

The \GCel\ uses T805 transputer CPU's at a clockspeed of $30\,$MHz. Each
transputer has a peak performance of $4.3\,$MFlops and a {\tt LINPACK
100x100} performance of $0.63\,$MFlops. The \GCel\ of the \PC2\ has
$4\,$MB memory per node. The communication between processors has a
sustained performance of $8.8\,$MB/s. The \GCel\ runs under {\tt PARIX
1.2} and uses the {\tt ACE} Fortran compiler and \PC2's version of the
\MPI\ libraries.

\section{Parallel radiative transfer}

\subsection{Strategy and implementation}

We use the method discussed in \cite{phhs392} for the numerical
solution of the special relativistic radiative transfer equation (RTE)
at every wavelength point \cite[see also][]{hsb94}.  This
iterative scheme is based on the operator splitting approach. The  RTE
is written in its characteristic form and the formal solution along the
characteristics is done using a piecewise parabolic integration \cite[PPM,
this is the ``short characteristic method''][]{ok87}.  We use the exact
band-matrix subset of the discretized $\Lambda$-operator as the
'approximate $\Lambda$-operator' (ALO) in the operator splitting
iteration scheme \cite[see][]{hsb94}. This has the advantage of
giving very good convergence and high speed-up when compared to diagonal
ALO's.

 The serial \RT\ code has been optimized for superscalar and 
vector computers and is numerically very efficient. 
It is therefore crucial to optimize the ratio of communication to 
computation in the parallel implementation of the \RT\ method. In terms
of CPU time, the most costly parts of the \RTC\ are the setup of the 
PPM interpolation coefficients and the formal solutions (which have
to be performed in every iteration). The construction of a tri-diagonal
ALO requires about the same CPU time as a single formal solution
of the RTE and is thus not a very important contributor to the total
CPU time required to solve the RTE at every given wavelength point.

In principle, the computation of the PPM coefficients does not require
any communication and thus could be distributed arbitrarily between
the nodes. However, the formal solution is recursive along each
characteristic. Within the formal solution, communication is only
required during the computation of the mean intensities $J$, as they
involve integrals over the angle $\mu$ at every radial point. Thus, a
straightforward and efficient way to parallelize the \RTC\ is to
distribute sets of characteristics onto different nodes.  This will
minimize the communication during the iterations, and thus optimize
the performance. Within one iteration step, the current values of the
mean intensities need to be broadcast to all \RT\ nodes and the new
contributions of every \RT\ node to the mean intensities at every
radius must be sent to the master node. The master \RT\ node then
computes and broadcasts an updated $J$ vector using the operator splitting scheme,
the next iteration begins, and the process continues until the solution
is converged to the required accuracy. The setup, i.e., the
computation of the PPM interpolation coefficients and the construction
of the ALO, can be parallelized using the same method and node
distribution. The communication overhead for the setup is roughly
equal to the communication required for a single iteration.

 An important point to consider is the load balancing between the
$\nrt$ \RT\ nodes. The workload to compute the formal solution along
each characteristic is proportional to the number of intersection points
of the characteristic with the concentric spherical shells of the radial
grid (the `number of points' along each characteristic). Therefore, if
the total number of points is $N_{\rm P}$, the optimum solution would be
to let each \RT\ node work on $N_{\rm P}/\nrt$ points. This optimum can,
in general, not be reached exactly because it would require splitting
characteristics between nodes (which involves both communication and
synchronization). A simple load distribution based on $N_{\rm
C}/\nrt$, where $N_{\rm C}$ is the total number of characteristics, is
far from optimal because the characteristics do not have the same number
of intersection points (consider tangential characteristics with
different impact parameters).  We therefore chose a compromise of
distributing the characteristics to the \RT\ nodes so that the total
number of points that are calculated by each node is roughly the same
and that every node works on a different set of characteristics.

\subsection{Performance and scalability}

 Following the outline given in the previous paragraphs, we have adapted
the serial version of our \RTC\ using the \MPI\ libraries.  The
additions required to implement the parallel version were relatively
small and required only the addition of the \MPI\ subroutine calls.
The total number of \MPI\ statements in the \RTC\ is 220, which is
very small compared to a total of 10,700 statements.

 We test the performance and scalability of the parallel \RTC\ with a
simple single wavelength \RT\ model on an \SP2. The test model uses 128
radial points, we have chosen the following test parameters: the radial
extension is a factor of 10, the total optical depth is 100, the ratio
of absorptive to total opacity is $\epsilon=10^{-4}$, and we require as
convergence criterion that changes in the mean intensities are less than
$10^{-8}$ between consecutive iterations (which requires 17 iterations
starting with $J=B$). For simplicity, a static model is used, the tests
of the full \phoenix\ code described below will employ test models with
velocity fields. In table \ref{s3-taba} we list
the results of test runs with different number of nodes.

 The scaling of the \RTC\ is acceptable as the number
of nodes rises  to about 8 nodes.
More nodes do not decrease the wall-clock time significantly because the
rapidly increasing communication overhead (the $J$ must be broadcast to
all nodes and the results must be gathered from all nodes in every \RT\
iteration) as well as the worsening of the load balance (the
latter becomes more acute as the number of nodes increases to $\ge 16$). 
We have verified that the individual modules of the \RTC\
(i.e., PPM coefficients, ALO computation, and formal solution) indeed
scale with the number of nodes, within the limits of the load balancing
described above. In test calculations we found that the
time for either the gathering of the results of the formal solution
(using {\tt MPI\_REDUCE}) or the broadcast of the updated $J$ (using
{\tt MPI\_BCAST}) {\em individually} do not require significant amount
of time. However, the {\em combination} of these two communications use
much more time than would be predicted by the sum of the times for the
individual operations. This is probably partly due to the required
synchronization but also partly due to limitations of the \MPI\
implementation and the communication hardware of the \SP2.

 Table \ref{s3-taba} also shows the results obtained on the 
\GCpp\ for the same test case for comparison. On a single CPU, the
\GCpp\ is about a factor of 4.5 slower than a single CPU of the \SP2,
which is consistent with their {\tt LINPACK} results. The scaling 
of the \GCpp\ with more nodes is about the same as for the \SP2, 
indicating that the slower communication speed of the \GCpp\ does
not produce performance problems if the number of CPU's is small. 
In Table \ref{s3-taba} we also include the results for the \HPJ. 
The speed-up for 2 CPU's is only 30\%, which is due to the 
slower communication caused by a non-optimized \MPI\ library. 

 The \RT\ test problem with 128 radial grid-points was too large to fit
into the memory of a single transputer node of the \GCel\ without major
changes in the code (which would be very time consuming because of the
lack of a Fortran 90 compiler for both {\tt Parsytec} systems).
Therefore, we have also run a smaller test case with 50 radial
gridpoints.  The results are also listed in Table \ref{s3-tabb}. The
performance ratio of the results on a single CPU is, as in the previous
case, comparable to the ratio of the {\tt LINPACK} results: The code
runs about a factor of 31 faster on a single {\tt PPC 601} CPU than on
the {\tt T805} transputer.  However, there are now major differences in
the scalability between the two systems.  Whereas the scaling of the
\GCel\ results are slightly better than the results obtained for the
\SP2\ with the large test case, the wall-clock times do not scale well
for both the \SP2\ and the \GCpp, in contrast to its behavior in the large
test case.  The reason is the relatively slower communication (compared
to raw processor speed) of the \SP2 and the \GCpp: in the large test
case the floating point operations dominate over the communication. For
the \GCel\ with lower floating point performance but faster
communication than the other two machines, the scaling to more
processors is much better and comparable to the large test case for the
\SP2\ and \GCpp. This demonstrates that flexibility of the load
distribution is very important in order to obtain good performance on a
number of different machines.

\section{Parallelization of the line opacity calculation}


The contribution of spectral lines by both atoms and molecules is
calculated in \phoenix\ by direct summation of all contributing lines.
Each line profile is computed individually at every radial point for
each line within a search window (typically, a few hundred to few
thousand Doppler widths or about $1000\ang$). This method is more
accurate than methods that rely on pre-computed line opacity tables
(the ``opacity sampling method'') or methods based on distribution
functions (the ``opacity distribution function (ODF) method''). Both
ODF and the opacity sampling method
neglect the details (e.g., depth dependence) of the 
individual line profiles. This introduces systematic errors in the
line opacity because the pressures on the top and the bottom of the
line forming region are vastly different (several orders of magnitude
in cool dwarf stars) which causes very different pressure damping and
therefore differing line widths over the line forming region.  In
addition, the pre-computed tables require a specified and fixed
wavelength grid, which is too restrictive for NLTE calculations that
include important background line opacities.

 In typical model calculations we find that about 1,000 to 10,000
spectral lines contribute to the line opacity (e.g., in M dwarf model
calculations) at any wavelength point.  Therefore, we need to calculate
a large number of individual Voigt profiles at every wavelength point.
The subroutines for these computations can easily be vectorized and we
use a block algorithm with caches and direct access scratch files for
the line data to minimize storage requirements. A block is the number
of lines stored
in active memory, and the cache is the total number of blocks stored in
memory. When the memory size is exceeded, the blocks are written to
direct access files on disk. Thus the number of
lines that can be included is not limited by RAM, but rather by disk
space and the cost of I/O. This approach is computationally efficient
because it provides high data and code locality and thus minimizes
cache/TLB and page faults. This has proven to be so effective that model
calculations with more than 15 million lines can be performed on medium
sized workstations (e.g., {\tt IBM RS/6000-530} with 64MB RAM and 360MB
scratch disk space). The CPU time for a single iteration for a LTE model
is dominated by calculating the line opacity (50 to 90\% depending on
the model parameters), therefore, both the LTE atomic and 
molecular line opacities are good candidates for parallelization.

There are several obvious ways to parallelize the line opacity
calculations.  The first method is to let each node compute the opacity
at $N_{\rm r}/\nnode$ radial points, the second is to let each node 
work on a different subset of spectral lines within each search window.
A third way, related to the second method, is to use completely
different sets of lines for each node (i.e., use a global workload split
between the nodes in contrast to a local split in the second method).
In the following, we will discuss the advantages and disadvantages of
these three methods. All three methods require only a very small 
amount of communication, namely a gather of all results to the
master node with an {\tt MPI\_REDUCE} library call.

 The first and second methods, distributing sets of radial points or
sets of lines within the local (wavelength dependent) search window
over the nodes, respectively, are very simple to implement. They can
easily be combined to optimize their performance: if for any
wavelength point the number of depth points is larger than the number
of lines within the local search window, then it is more effective to
run this wavelength point parallel with respect to the radial points,
otherwise it is better to parallelize with respect to the lines within
the local window. It is trivial to add logic to decide the optimum
method for every wavelength point individually. This optimizes overall
performance with negligible overhead. The speed-up for this method of
parallelization is very close to the optimum value if the total number
of blocks of the blocking algorithm is relatively small ($\le 3\ldots
5$).

 However, if the total number of line blocks is larger (typically,
about 10 to 20 blocks are used), the overhead due to the read operations
for the block scratch files becomes noticeable and can reach
20\% or more of the total wall-clock time. This is due to the fact that
the `local parallelization' required that each node working on the
line opacities needs to read every line block, thus increasing the
IO time and load to the IO subsystem by the number of nodes themselves.

An implementation of the third method, i.e., distributing
the global set of lines onto the nodes, will  therefore be more effective if
the number of lines is large (this is the usual case in 
practical applications). There are a number of ways to implement this
approach; however, many of them would require either significant
administration and communication during the line selection procedure
or a large number of individual scratch files (one for each node).
The by far easiest and fastest way to implement the global 
distribution on the \SP2\ is to use the Parallel IO Filesystem 
(\piofs). The \piofs\ has the advantage that (a) the code changes
required are simple and easily reversible (for compatibility with 
other machines), (b) the \piofs\ software allows the creation of 
a single file on the line selection nodes (the line selection is
a process that is inherently serial and can be parallelized only
by separating the atomic and molecular line selections, which although
simple has not yet been implemented), (c) it allows different nodes to access distinct
portions of a single file as their own `subfiles', and (d) that the
IO load is distributed over all IO nodes that are \piofs\ servers.

 Points (b) and (c) make the implementation very simple. We chose
a single line
as the atomic size of the \piofs\ file that is used to store the 
line-blocks and create the \piofs\ file with
$\nnode$ sub-files (where $\nnode$ is the number of nodes that 
will later work on the LTE line opacity). The line selection
routines then set the \piofs\ file view to the equivalent of a 
single direct access file with the appropriate block size. After the
line selection, each of the $\nnode$ nodes sets the local
view of the \piofs\ file such that it `sees' the $1 \ldots \nnode$
subfile, so that node $N$ reads lines $N$, $N+\nnode$, $N+2\nnode$, 
and so forth. This is equivalent to the `global line distribution'
method with a minimal programming effort. The advantages are
not only that each node has to read only $1/\nnode$ of the total
lines but also that the IO itself is distributed both over the
available \piofs\ fileservers and over time (because the different
line sets will cause the block IO operations to happen at different
times). The latter is very useful and enhances the parallel performance 
and scalability. In addition, the process is completely transparent
to the program, only 2 explicit \piofs\ subroutine calls
had to be inserted to set the `view' of the \piofs\ file to the
correct value.  

\begin{figure}[ht]
\begin{center}
\leavevmode
\psfig{file=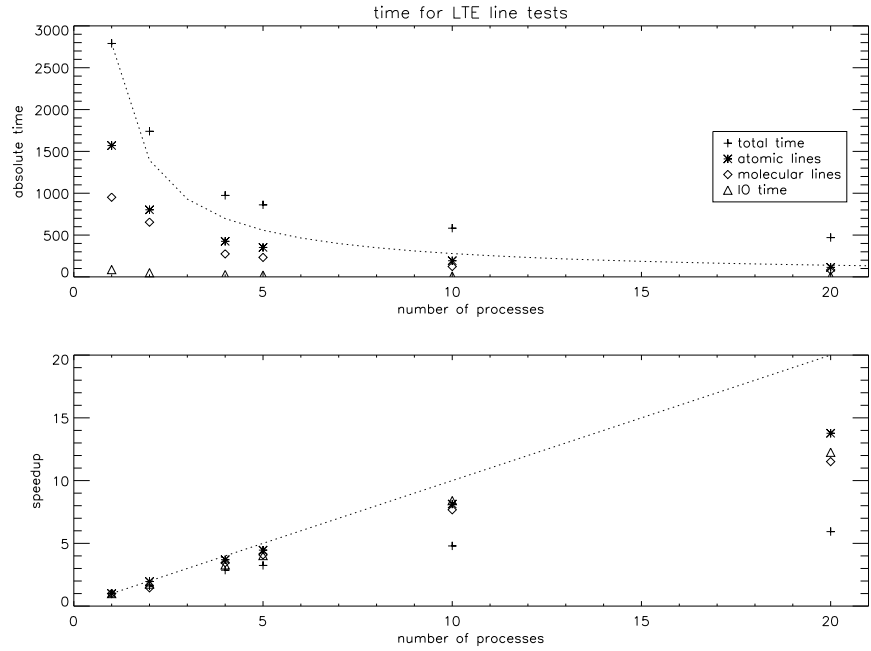,width=12cm,angle=90}
\caption[]{\label{lte-lines}The speed-up and scaling of the line
opacity calculation using \piofs\ for 
different numbers of \SP2\ nodes used.
The upper panel with the 
wall-clock time (in seconds) used for the test calculations, the lower
panel the speed-up factors. The dotted curves give the expected maximum
parallel speed-up.}
\end{center}
\end{figure}

 In Fig.~\ref{lte-lines} and Table \ref{ltelines-tab}
we show the performance and speed-up 
as function of the number of nodes for a simplified model
of a very low mass star. 
We use an M dwarf star model with the parameters $\Teff=2700\k$,
$\logg=5.0$, and solar abundances, appropriate for the late
M dwarf VB10 \cite[]{svb10pap}. The test model includes 226,000 atomic
lines (of these, 29,000 are calculated using Voigt profiles)
and 11.2 million molecular lines (with 3.6 million calculated
with Voigt profiles) and uses a wavelength grid with 13,000 points.
We have set the line search windows to values larger than
required in order to simulate ``real'' model calculations
(which typically use twice as many wavelength points). In
addition, we set the blocksizes
for the line data storage to 30,000 lines for both molecular
and atomic lines and used a line data cache size of 2 (that is we
store in RAM 2 blocks of 30,000 lines each). This setting
is optimal for the atomic lines but the blocksizes are 
set to larger values (about 100,000) for the molecular lines
in production models. However, the cache sizes were large 
enough to prevent cache thrashing. 

 The speed-up that we find is very good and the scaling is
close to the theoretical maximum for the atomic lines (a factor
of 4.5 for 5 nodes). For the molecular lines the speed-up is
4.1 for 5 nodes, a little lower than for the atomic lines. This
is caused by the additional IO time required to read the 
line data blocks from the scratch file. Ignoring the IO time,
the molecular line speed-up is 4.6, very close to the value
found for the atomic lines. These results were obtained by using
the parallel IO system, if we instead rely on a standard file,
the IO time for the molecular lines increases by a factor of 3.2
and the speed-up for the molecular lines decreases to a factor
of only 2.2 (for 5 nodes, respectively). In production runs
on machines that do not have a \piofs\ filesystem, we would of course
use larger blocksizes for the line data arrays to minimize
the IO time, we usually do this on serial machines. 

\section{Parallelizing NLTE calculations}


Our method for iteratively solving the NLTE radiative transfer and
rate equations with an operator splitting method are discussed in
detail in \cite{phhcas93}~and~\cite{hbfe295}, therefore, we present here only
a brief summary. The method uses a ``rate-operator'' formalism that
extends the approach of \cite{rybhum91} to the general
case of multi-level NLTE calculations with overlapping lines and
continua.  We use an ``approximate rate-operator'' that is constructed
using the exact elements of the discretized $\Lambda$-matrix (these
are constructed in the radiative transfer calculation for every
wavelength point). This approximate rate-operator can be either
diagonal or tri-diagonal. The method gives good convergence for a wide
range of applications and is very stable. It has the additional
advantage that it can handle very large model atoms, e.g., we use a
617 level Fe~II model atom in regular model calculations
\cite[]{phhnovfe296,snefe296}.

 Parallelizing the NLTE calculations involves parallelizing three
different sections: the NLTE opacities, the rates, and the solution of
the rate equations. In the following discussion, we consider only a
diagonal approximate rate-operator for simplicity. In this case, the
computation of the rates and the solution of the rate equations as
well as the NLTE opacity calculations can simply be distributed onto a
set of nodes without any communication (besides the gathering of the
data at the end of the iteration). This provides a very simple way of
achieving parallelism and minimizes the total communication
overhead. The generalization to a tridiagonal approximate rate-operator
is in principle straightforward, and involves only communication at the
boundaries between two adjacent nodes.

 It would be possible to use the other two methods that we discussed
in the section on the LTE line opacities, namely local and global
set of lines distributed to different nodes. However, both would 
involve an enormous amount of communication between the NLTE nodes because
each NLTE transition can be coupled to any other NLTE transition. These
couplings require that a node working on any NLTE transition have the 
required data to incorporate the coupling correctly. Although this
only applies to the nodes that work on the NLTE rates, it would require
both communication of each NLTE opacity task with each other
(to prepare the necessary data) and communication from the NLTE 
opacity nodes to the NLTE rate nodes. This could mean that several
MB data would have to be transferred between nodes at each
wavelength point, which is prohibitive with current
communication devices.

 Another way to parallelize the NLTE calculation exploits the fact that
our numerical method allows the grouping of NLTE species (elements) into
separate entities. These groups do not need to communicate with each
other until the end of an iteration. Thus distributing groups onto
different nodes will allow very effective parallelization with little or
no communication overhead. This approach can be combined with the all
the parallelization approaches discussed previously, leading to better
speed-ups. In addition, it would significantly reduce the memory
requirements for each node, thus allowing even larger problems to be
solved. We will implement this method in later releases of \phoenix\ and
report the results elsewhere.

 We have implemented the parallelization of the 
NLTE calculations by distributing the set of radial points on different
nodes. In order to minimize communication, we also `pair' NLTE nodes
so that each node works on NLTE opacities, rates and rate equations
for a given set of radial points. This means that the communication at
every wavelength point involves only gathering the NLTE opacity
data to the \RT\ nodes and the broadcast of the result of the 
\RT\ calculations (i.e., the ALO and the $J$'s) to the nodes 
computing the NLTE rates. The overhead for these operations is negligible
but it involves synchronization (the rates can only be computed after
the results of the \RT\ are known, similarly, the \RT\ nodes have to
wait until the NLTE opacities have been computed). Therefore, a good 
balance between the \RT\ tasks and the NLTE tasks is important
to minimize waiting. The rate equation calculations parallelize
trivially over the layers and involve no communication if the diagonal
approximate rate-operator is used. After the new solution
has been computed, the data must be gathered and broadcast to {\em all}
nodes, the time for this operation is negligible because it is 
required only once per model iteration.

\begin{figure}[ht]
\begin{center}
\leavevmode
\psfig{file=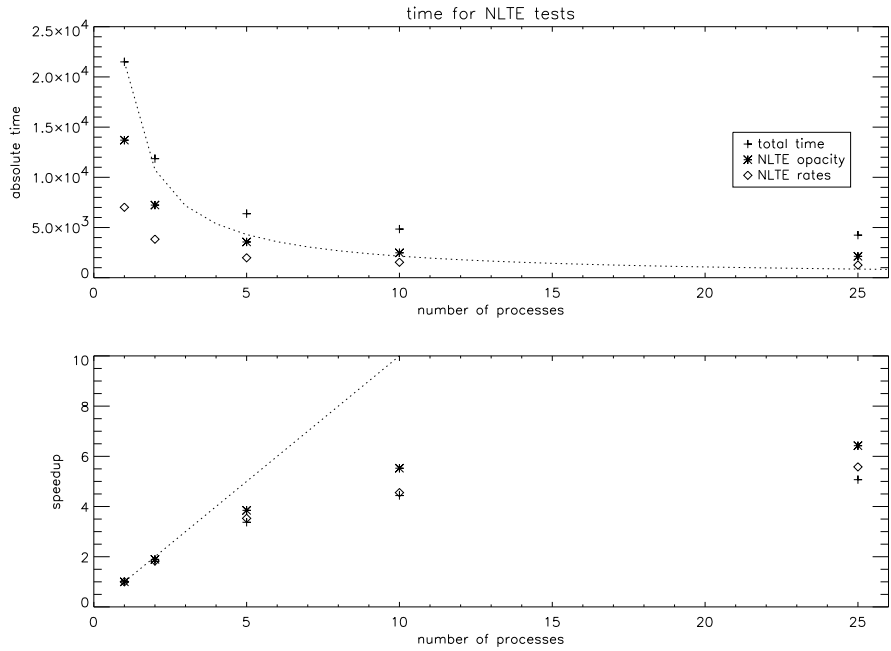,width=12cm,angle=90}
\caption[]{\label{dz-opt}The speed-up and scaling of the NLTE parallel
algorithm using Fortran-90 {\tt MODULE}s and individual array allocation
for different numbers of \SP2\ nodes used. The upper panel with the
wall-clock time (in seconds) used for the test calculations, the lower
panel the speed-up factors. The dotted curves give the expected maximum
parallel speed-up}
\end{center}
\end{figure}

In Fig.~\ref{dz-opt} and Table \ref{dz-tab} 
we show performance results for a simplified
test model with the following parameters: $\Teff=20,000\k$,
$\logg=8.0$, and solar abundances. The background
LTE line opacities have been omitted and the \RTC\ has been
run in plane parallel mode on a single node to concentrate the results
on the NLTE calculation. The speed-ups are acceptable although they
do not reach the theoretical maximum. This is caused by several
effects. First, the NLTE opacity calculations involve the computation
of the b-f cross-section, which has not been parallelized in this
version of the code. These calculations are negligible in serial mode,
but they become comparatively more costly in parallel mode. The solution
to this will be their parallelization, which involves significant
communication at each wavelength point. This will be investigated in 
future work. 

An important problem arises from the fact that the time spent in
the NLTE routines is not dominated by the floating point operations
(both the number and placement of floating point operation were
optimized in the serial version of the code) but by {\em memory access},
in particular for the NLTE rate construction. Although the parallel
version accesses a much smaller number of storage cells (which naturally
reduces the wall-clock time), effects like cache and TLB misses and page
faults contribute significantly to the total wall-clock time. All of
these can be reduced by using Fortran-90 specific constructs in the
following way: We have replaced the static allocation of the arrays that
store the profiles, rates etc.\ used in the NLTE calculations (done with
{\tt COMMON} blocks) with a Fortran-90 module and explicit allocation
(using {\tt ALLOCATE} and {\tt SAVE}) of these arrays at the start of
the model run. This allows us to tailor the size of the arrays to fit
exactly the number of radial points handled by each individual node.
This reduces both the storage requirements of the code on every node
and, more importantly, it minimizes cache/TLB misses and page faults. In
addition, it allows much larger calculations to be performed because the
RAM, virtual memory and scratch disk space of every node can be fully
utilized, thus effectively increasing the size of the possible
calculation by the number of nodes.

We find that the use of adapted array sizes significantly improves the
overall performance, even for a small number of processes. The scaling
with more nodes has also improved considerably. However, the overhead of
loops cannot be reduced and thus the speed-up cannot increase
significantly if more nodes are used. We have verified with a simple
test program which only included one of the loops important for the
radiative rate computation that this is indeed the case. Therefore, we
conclude that further improvements cannot be obtained at the Fortran-90
source level but would require either re-coding of routines in assembly
language (which is not practical) or improvements of the
compiler/linker/library system. We note that these performance changes
are very system dependent. For example, the serial ``{\tt COMMON} block''
version of the code runs significantly (50\%) faster than the serial
``Fortran-90 module'' version on both Cray's or SGI's, whereas on {\tt
IBM} {\tt RS/6000}'s we found no noticeable difference in speed. This
stresses that raw CPU performance is irrelevant as long as it is not
supplemented by adequate compilers.

 We also ran tests on the \HPJ\ machine, see Table \ref{dz-tab}. We use
the public domain {\tt MPICH} implementation of \MPI\ compiled on this
machine with its default compiler options.  There is no full-F90
compiler available from Hewlett-Packard for this machine, therefore, we
had to use the ``{\tt COMMON} block'' version of the code.  Therefore,
the speed-ups cannot be expected to be optimal. However, a total
speed-up of about 1.5 is acceptable. The speed-up for the opacity part of
the NLTE calculation is a factor of 1.7, somewhat better than the factor
of 1.4 achieved for the NLTE rate calculations. This is probably due to
the fact that the rate computation is much more memory access dominated
than the more floating point intensive NLTE opacity calculations.

\section{Performance for realistic full model calculations}

\subsection{Supernovae}
\begin{figure}[ht]
\begin{center}
\leavevmode
\psfig{file=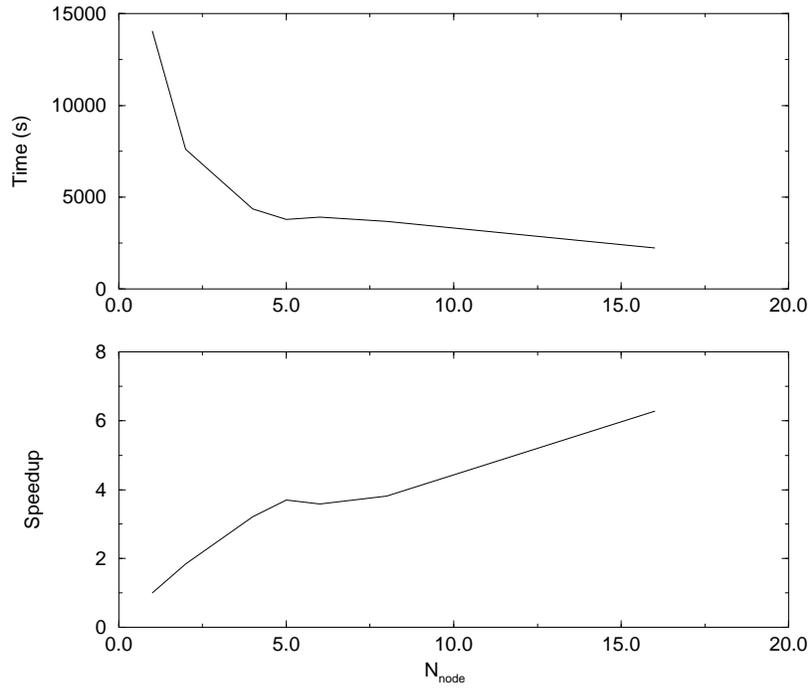,width=12cm,angle=-90}
\caption[]{\label{sn1a}The speed-up and scaling of the NLTE parallel
algorithm for a practical application
for different numbers of \SP2\ nodes used. The upper panel with the
wall-clock time (in seconds) used for the calculations, the lower
panel the speed-up factors.}
\end{center}
\end{figure}

In order to test the performance of the parallel code in a practical
application with velocity fields we calculated a supernova atmosphere
model with parameters typical for a Type Ia supernova explosion. The
total number of wavelength points = 48397, the total number of levels
treated in NLTE = 3752, and the total number of primary NLTE
transitions = 16594, the number of secondary NLTE transitions =
184943, and the number of LTE lines=2346474. Fig.~\ref{sn1a}a displays
the wall-clock time for a single model iteration as a function of
processors and Fig.~\ref{sn1a}b shows the speed-up, which in agreement
with our above results scales roughly as $\nnode/2$. While this is
significantly below the theoretical maximum of $\nnode$ scaling, it is a
significant speed-up for practical applications.

\subsection{Central stars of nova systems}
 To investigate the performance gains for a realistic 
stellar model at higher temperatures, we ran a complete NLTE
model calculation for a set of parameters expected for the 
primary component of a nova system in a quiet post-outburst 
phase. The parameters appropriate for these white dwarf systems
are $\Teff=21,000\k$, $\logg=8.0$ and solar abundances. We
use the following NLTE species:  H~I--II, He~I--III, C~I--IV,
N~I--VI, O~I--VI, Mg~II, Ca~II, S~II--II, Si~II-III, Ne~I,
and Fe~II with a total of 2,826 NLTE levels and 26,874 NLTE lines.
All NLTE lines are treated with detailed Voigt profiles. In addition,
we include 621,920 LTE background lines and the calculation uses a total of
93,619 wavelengths points. On a {\tt SGI} {\tt Power} {\tt Indigo 2}
a single model iteration with this setup requires about 9.5h CPU time
and the full model run (10 iterations) needs about 2 days CPU time.
On 10 processors of the \SP2\ (using a standard load-distribution
not optimized for the model parameters), a single iteration takes
about 1.4h wall-clock time and the full model calculation needs
about 7.3h wall-clock time, which is {\em less} than the CPU time
for a {\em single} iteration on the {\tt SGI} {\tt Power} {\tt Indigo 2}.
This shows that the parallel speed-ups that we can achieve in realistic
calculations are very significant, even with a small number of CPU's.
With about 7.5h wall-clock time for a complete model iteration,
substantial grids of these models can be constructed in relatively
short time, thus making detailed analysis of observed spectra 
with the best input physics feasible.

\subsection{M dwarfs}
 We have run a realistic NLTE M~dwarf test calculation with the
 following model parameters: $\Teff=2700\k$, $\logg=5.0$, and solar
abundances. As NLTE species we include H~I (10 levels), Na~I (3 levels),
Ti~I (395 levels), Ti~II (204 levels), C~I (228 levels), C~II (85
levels), N~I (252 levels), N~II (152 levels), O~I (36 levels), and O~II
(171 levels), for a total of 1591 levels and 15062 primary NLTE lines
(treated in detail with individual Voigt profiles). We use 113,433
wavelength points and include 288,775 background atomic LTE lines (with
28,011 of these are strong enough to be included with individual Voigt
profiles) as well as 12,861,979 molecular LTE lines (including 3,753,353
with Voigt profiles). 

In Table \ref{mdwarf-nlte-tab} we give the
wall-clock times for one iteration on the \SP2. We have used a ``small''
code configuration with blocksizes appropriate for machines with about
128MB RAM per node although the test machine had up to 300 per node
paging space
and we used very large search windows for the atomic, molecular and NLTE
lines in order to obtain a ``worst case'' scenario. Table
\ref{mdwarf-nlte-tab} shows that the calculation is dominated by the LTE
atomic and molecular line opacity whereas the NLTE opacities and rates
are only a second order contribution to the total time per iteration.
The scaling of the calculation is, therefore, very good up to the
largest configuration that we have tested. We could not run
the test model on a single \SP2\ CPU due to both wall-clock
time and memory restrictions, this demonstrates the importance
of parallelization for practical applications.

There are possibilities to
reduce the wall-clock time by, e.g., using larger blocksizes and a
specially tuned load-distribution. The last \SP2\ entry in Table
\ref{mdwarf-nlte-tab} shows that an alternative load distribution can
easily improve the overall speed although now some of the sub-tasks require
more wall-clock time.  

We also include the timing results of the test run that we obtained
on a single processor of a Cray C90 (CPU times). The Table shows that
the C90 is about as fast as 5 nodes of the \SP2, which is roughly
the relative performance ratio of a single \SP2\ node to a single
C90 processor. The wall-clock time on the C90 was much worse
than on the \SP2, due to the time-sharing operation of the C90 CPUs.

\section{Conclusions}

We have been able to obtain a significant speed-up of our serial model
atmosphere code with only a modest number of changes. We find that using
the \MPI\ library calls and a distributed memory model the parallel
version of \phoenix\ was easier to add to our existing serial code than
a shared memory model where we have to carefully make sure that specific
memory locations are only updated correctly. The speed-ups we have been
able to achieve are below the theoretical maximum, however, this is not
unexpected when such things as loop overhead and communications are
accounted for. This is very similar to the earlier process of moving
from strictly serial codes to vectorized codes, the theoretical maximum
vector speed-up is very rarely reached in practical applications.  The
parallel speed-up of \phoenix\ is important for practical application
and, in addition, allows both much larger (in terms of memory size and
CPU time) problems to be handled.

 Future improvements of the parallel version of \phoenix\ will
include the distribution of the NLTE groups to different nodes
(improving the degree of parallelization and allowing much 
larger problems to be handled on machines with less memory per
node, e.g, the Cray {\tt T3D}) as well as additional optimization
of the code based on experience with large scale production
runs on parallel machines. 

%

\acknowledgments
It is a pleasure to thank D. Branch, P. Nugent, A. Schweitzer,
S. Shore, and S. Starrfield for stimulating discussions. We thank the
anonymous referee for suggestions which improved the presentation.
This work was supported in part by NASA LTSA grants NAGW 4510 and NAGW
2628 and by NASA ATP grant NAG 5-3067 to Arizona State University and
by NSF grant AST-9417242, and by an IBM SUR Grant to the University of
Oklahoma. Some of the calculations presented in this paper were
performed at the Cornell Theory Center (CTC), the San Diego
Supercomputer Center (SDSC), supported by the NSF, and the Paderborn
Center for Parallel Computing,  we thank them for a
generous allocation of computer time.


\clearpage


\begin{deluxetable}{rrrr}
\tablecolumns{4}
\tablenum{1a}
\tablewidth{0pc}
\tablecaption{\label{s3-taba}Results of radiative transfer tests.}
\tablehead{\multicolumn{4}{c}{Test with 128 radial points}\\
\colhead{nodes} &  \colhead{\SP2} & \colhead{\GCpp} &
\colhead{\tt HP J200}
}
\startdata
     1  &  306.4 : 1.0 & 1410.6 : 1.0 & 818.8 : 1.0 \\
     2  &  159.7 : 1.9 & 776.4 : 1.8 & 652.8 : 1.3 \\
     3  &  108.0 : 2.8 & 590.2 : 2.4 \\
     4  &  94.8 : 3.2 & 512.1 : 2.8 \\
     8  &  63.8 : 4.8 \\
    16  &  53.3 : 5.7 \\
    32  &  56.1 : 5.5 \\
    64  &  68.1 : 4.5 \\
\enddata
\tablecomments{The results are given in the format $t\ :\ x$ where $t$ is the absolute wall-clock 
time in seconds and $x$ is the speed-up factor relative to the
serial run, rounded to 2 significant figures.}
\end{deluxetable}

\begin{deluxetable}{rrrr}
\tablecolumns{4}
\tablenum{1b}
\tablewidth{0pc}
\tablecaption{\label{s3-tabb}Results of radiative transfer tests.}
\tablehead{\multicolumn{4}{c}{Test with 50 radial points}\\
\colhead{nodes} &  \colhead{\SP2} & \colhead{\GCel} & \colhead{\GCpp} 
}
\startdata
1 & 72.8 : 1.0 & 5057.0 : 1.0 & 164.1 : 1.0\\
2 & 45.1 : 1.6 & 2473.1 : 2.0 & 117.1 : 1.4\\
4 & 33.0 : 2.2 & 1447.0 : 3.5 & 107.0 : 1.5\\
8 & 31.1 : 2.3 & 959.9 : 5.3 \\
16 & 34.5 : 2.1 & 754.8 : 6.7 \\
32 & 42.6 : 1.7 & 741.5 : 6.8 \\
64 & 62.9 : 1.2 & 907.7 : 5.6 \\
\enddata
\end{deluxetable}

\begin{deluxetable}{rrrrr}
\tablenum{2}
\tablewidth{0pc}
\tablecaption{\label{ltelines-tab}Results of the line routine 
parallelization tests.}
\tablehead{\colhead{nodes} &  \colhead{atomic lines} &
\colhead{molecular lines} 
& \colhead{total time} & \colhead{IO time}} 
\startdata
& \multicolumn{4}{c}{\SP2\ \piofs}\\
   1 & 1570.0 : 1.0 & 952.0 : 1.0 & 2790.0 : 1.0 & 87.4 : 1.0 \\
   2 & 802.0 : 2.0 & 654.0 : 1.5 & 1740.0 : 1.6 & 49.8 : 1.8 \\
$*$2 & 769.0 : 2.0 & 539.0 : 1.8 & 1520.0 : 1.8 & 73.9 : 1.2 \\
   4 & 425.0 : 3.7 & 275.0 : 3.5 & 974.0 : 2.9 & 26.6 : 3.3 \\
$*$4 & 388.0 : 4.0 & 304.0 : 3.1 & 899.0 : 3.1 & 67.2 : 1.3 \\
   5 & 352.0 : 4.5 & 233.0 : 4.1 & 861.0 : 3.2 & 21.7 : 4.0 \\
  10 & 193.0 : 8.2 & 124.0 : 7.7 & 582.0 : 4.8 & 10.4 : 8.3 \\
  20 & 114.0 : 13.8 & 82.7 : 11.5 & 470.0 : 5.9 & 7.1 : 12.2 \\
\hline
 & \multicolumn{4}{c}{\SP2\ standard} \\
 5 & 331.0 : 4.8 & 442.0 : 2.2 & 1170.0 : 2.4 & 206.0 : 0.4 \\
\hline
 & \multicolumn{4}{c}{Cray C90} \\
   1 & 276.0      & 175.0      & 535.0 \\
\enddata
\tablecomments{The results
are given in the format $t\ :\ x$ where $t$ is the absolute wall-clock 
time in seconds and $x$ is the speed-up factor relative to the
serial run. We include also a run done without using the \piofs\ 
filesystem.The entries marked with ``$*$'' have been obtained on
the \SP2\ of the University of Oklahoma, which has more RAM per node
than the CTC machine. The longer I/O time on the OU machine may be
related to the fact that both \piofs\ and NFS mounts use the high
speed switch on this machine.}
\end{deluxetable}

%

\begin{deluxetable}{rrrr}
\tablenum{3}
\tablewidth{0pc}
\tablecaption{\label{dz-tab}Results of the NLTE routine 
parallelization tests.}
\tablehead{\colhead{nodes} &  \colhead{NLTE opacity} & \colhead{NLTE
rates} & \colhead{total}}
\startdata
 & \multicolumn{3}{c}{\SP2} \\
 1 & 13701.0 : 1.0 & 7015.3 : 1.0 & 21497.0 : 1.0  \\
 2 & 7234.3 : 1.9 & 3834.2 : 1.8 & 11860.0 : 1.8  \\
 5 & 3562.0 : 3.8 & 1987.2 : 3.5 & 6376.3 : 3.4  \\
10 & 2479.4 : 5.5 & 1540.4 : 4.6 & 4844.2 : 4.4  \\
25 & 2133.1 : 6.4 & 1257.6 : 5.6 & 4240.2 : 5.1  \\
\hline
 & \multicolumn{3}{c}{\tt HP J200} \\
 1 & 16102.0 : 1.0 & 11525.0 : 1.0 & 28554.0 : 1.0 \\
 2 & 9520.7 : 1.7 & 8458.4 : 1.4 & 19342.0 : 1.5 \\
\enddata
\tablecomments{The results
are given in the format $t\ :\ x$ where $t$ is the absolute wall-clock 
time in seconds and $x$ is the speed-up factor relative to the
serial run.}
\end{deluxetable}

\begin{deluxetable}{rrrrrr}
\tablecolumns{6}
\tablenum{4}
\tablewidth{0pc}
\tablecaption{\label{mdwarf-nlte-tab}Results of a M dwarf model
run including NLTE effects.}
\tablehead{\colhead{}       & \colhead{atomic} & \colhead{molecular} & \colhead{NLTE} & \colhead{NLTE} & \colhead{total}\\
\colhead{nodes} & \colhead{lines} & \colhead{lines} & \colhead{opacity} & \colhead{rates} & \colhead{time}}
\startdata
      & \multicolumn{5}{c}{\SP2} \\
    2 & 13900.0 & 6340.0& 2060.0 & 760.0 & 24900.0\\
    5 & 5740.0 & 2680.0 & 1100.0 & 519.0 & 11800.0\\
   10 & 3160.0 & 1300.0 & 912.0 & 505.0 & 7980.0\\
   20 & 1670.0 & 700.0 & 791.0 & 420.0 & 6140.0\\
$*$20 & 1530.0 & 638.0 & 1280.0 & 646.0 & 5560.0\\
\hline
      & \multicolumn{5}{c}{Cray C90} \\ 
    1 & 4900.0 & 2290.0 & 1270.0 & 943.0 & 10100.0\\
\enddata
\tablecomments{The results
are given are the absolute wall-clock 
time in seconds for the \SP2\ and CPU times in seconds
for the {\tt C90}.
The last table entry for the \SP2\ (marked with a ``$*$'') uses a
more optimized load distribution that separates tasks. The results
in a speed-up of 10\% compared to the `simple' load distribution.}
\end{deluxetable}

\clearpage

\section{Figures}

\end{document}